\documentclass{jfm}
\usepackage{graphicx}
\usepackage{epstopdf, epsfig}
\usepackage{mathptmx}
\usepackage{xcolor}
\usepackage{bm}
\usepackage{hyperref}
\usepackage{makecell}
\usepackage{upgreek}

\newcommand{\rmnum}[1]{\romannumeral #1}
\newcommand{\Rmnum}[1]{\expandafter\@slowromancap\romannumeral #1@}

\shorttitle{Self-generated magnetic field in 3D ARTI}
\shortauthor{D. Zhang et al.}

\title{Self-generated magnetic field in three-dimensional ablative Rayleigh-Taylor instability}

\author{Dehua Zhang\aff{1}, Xian Jiang\aff{1}, Tao Tao\aff{2}, Jun Li\aff{3}, Rui Yan\aff{1,4}
\corresp{\email{ruiyan@ustc.edu.cn}}, De-Jun Sun\aff{1},
\and Jian Zheng\aff{2,4}}

\affiliation{
\aff{1}Department of Modern Mechanics, University of Science and Technology of China, Hefei 230026, China
\aff{2}Department of Plasma Physics and Fusion Engineering, University of Science and Technology of China, Hefei, Anhui 230026, China
\aff{3}Institute of Applied Physics and Computational Mathematics, Beijing 100094, China
\aff{4}IFSA Collaborative Innovation Center, Shanghai Jiao Tong University, Shanghai 200240, China
}

\begin{document}

\maketitle

\begin{abstract}
The self-generated magnetic field in three-dimensional (3D) single-mode ablative Rayleigh-Taylor instabilities (ARTI)  relevant to the acceleration phase of a direct-drive inertial confinement fusion (ICF) implosion is investigated. It is found that stronger magnetic fields up to a few thousands of T can be generated by 3D ARTI than by its two-dimensional (2D) counterpart. The Nernst effects significantly alter the magnetic fields convection and amplify the magnetic fields. The scaling law for the magnetic flux obtained in the 2D simulations performs reasonably well in the 3D cases. While the  magnetic field significantly accelerates the bubble growth  in the  short-wavelength 2D modes through modifying the heat fluxes, the magnetic field mostly accelerates the spike growth but has little influence on the bubble growth in 3D ARTI.
\end{abstract}


\section{Introduction}\label{Introduction}
The Rayleigh-Taylor instability (RTI) \citep{Rayleigh1900, Taylor1950} is a fundamental hydrodynamic instability that occurs at the interface between heavy and light fluids when the heavy fluid is supported by the light fluid against gravity. RTI plays  important roles in a number of astrophysical processes such as supernova explosions \citep{Burrows2000,Gamezo2003}, and is considered as a critical risk in inertial confinement fusion (ICF) \citep{Lindl1998,Atzeni2004} implosions. In a typical ICF experiment, a cold spherical deuterium and tritium (DT) target is irradiated either by direct laser light in the direct-drive \citep{Craxton2015} approach, or by X-rays emitted by a high-Z hohlraum \citep{Lindl1998} in the indirect-drive \citep{Lindl1995,Lindl2004} approach. The energy absorbed near the target outer surface rapidly heats the material up and causes intense mass ablation off the target shell, leading to the shell's inward acceleration and compression. As the higher-density target shell is accelerated by the lower-density ablated plasma, the interface perturbed by initial surface roughness or irradiation non-uniformity is unstable to RTI which develops into an interchange of heavy with light fluids. The light fluid rises up forming ``bubbles'' while the heavy fluid falls down forming ``spikes''. RTI dramatically degrades the implosion performances by compromising the shell integrity and mixing the inside deuterium-tritium (DT) fuel with the outside high-Z ablator.  As the milestone on ignition has been recently achieved \citep{Zylstra2022,abu2022lawson} on the National Ignition Facility (NIF),  further improved implosion performance and higher gain are being pursued in the future ICF designs where controlling the hydrodynamic instabilities remains a key factor to be considered. 

As a candidate to improve ICF implosion performance, the applications of magnetic  fields in ICF have been attracting intensive research interests.
An externally applied magnetic field has been proposed as a promising approach to improve ICF implosion performance by reducing the electron thermal conduction and magnetically confining the DT-$\alpha$ burning plasma in the hot spot \citep{Wurden2016,Perkins2017}.  Enhanced fusion yield and temperature have been reported in cylindrical magnetized liner inertial fusion implosions \citep{Slutz2012,Gomez2014} and magnetized direct-drive ICF implosions \citep{Chang2011}. Recent experiments at NIF have also demonstrated performance enhancement from an applied magnetic field in room-temperature (“warm”) indirect-drive implosions \citep{Moody2022,Sio2023}. The effects of externally applied magnetic fields on RTI in ICF-relevant conditions were also studied via numerical simulations \citep{Walsh2017, Walsh2022} and experiments \citep{Matsuo2021}. \cite{Walsh2022} investigated the effects of externally applied external magnetic fields in different directions on the growth of the magnetized ARTI, considering both the magnetic tension and the magnetized heat flow, via three-dimensional (3D) extended-magnetohydrodynamics simulations.
The experimental work by \cite{Matsuo2021} found that the external magnetic field reduces the electron thermal conduction across the magnetic field lines and enhances the ARTI growth.

Another type of magnetic field is self-generated by the plasmas and companies with the evolution of hydrodynamic instabilities.
It was first predicted by the theoretical works that magnetic field can be generated in RTI in laser-produced plasmas \citep{Stamper1971,Mima1978,Haines1985,Stamper1991}. The Biermann-battery effect was identified as the key source generating magnetic field in RTI \citep{Mima1978} : The misaligned temperature and density gradients generate magnetic fields via the nonzero $\nabla T_e \times \nabla n_e$, where $T_e$ and $n_e$ are the electron temperature and electron number density, respectively. In experiments, simultaneous Faraday rotation was utilized to diagnose the spontaneous magnetic field in laser-produced plasmas \citep{Stamper1975,Raven1978,Stamper1978}, but such diagnostic techniques used external optical probing and thus were inadequate to measure inside the high-density plasmas opaque to the probing lights \citep{Wagner2004}. Proton radiography was widely used to diagnose the laser-driven magnetic field structures under more extreme plasma parameters \citep{Li2007,Li2009,Gao2012,Gao2013,Gao2015,Manuel2012,Manuel2015}. The proton radiography experiments have shown that Mega-Gauss(MG)-level magnetic fields can be generated in RTI in laser-produced plasmas \citep{Manuel2012,Manuel2015,Gao2013}.

Self-generated magnetic fields not only facilitate diagnostics on the deliberate fluid structures of RTI inside hot plasmas for the applications in proton radiography but may also influence hydrodynamic evolution if intense enough. While MG-level magnetic is not strong enough to directly affect the implosion hydrodynamics because the plasma thermal energy far exceeds the magnetic pressure (ie. the parameter $\beta \equiv 8\pi p/B^2 \gg 1$, where $p$ is the pressure) in ICF-relevant plasmas, it may be strong enough to magnetize the plasma and alter the electron thermal conduction when the cyclotron frequency of the electron reaches the same order of magnitude as the electron collision frequency. The importance of the magnetic modification on electron thermal conductivity is often evaluated by the Hall parameter \citep{Braginskii1965} $\chi=\omega_{ce}\tau_e$, where $\omega_e$ is the electron cyclotron frequency, $\tau_e$ is the characteristic time of electron collisions. The self-generated magnetic fields due to RTI in ICF is largely determined by the mass ablation feature which brings rich physics to not only the hydrodynamics but also  the generation and transportation of the magnetic fields.

When the intense laser energy is deposited on an ICF target shell, an ablated plasma outflow rapidly develops from the surface of shell (ablation front) and creates a high-temperature and low-density fluid relative to the unablated materials.
The laser-driven RTI is characterized by this ablation process on the outer surface of the shell during the acceleration phase of the implosion, and the ablative RTI (ARTI) consequently behaves quite differently from the classical RTI (CRTI) due to the mass ablation near the ablation front. 
Mass ablation can stabilize RTI in the linear phase and results in a linear cutoff wave number $k_{c}$ in the unstable spectrum \citep{Takabe1985,Betti1998} such that all modes with the perturbation wave number $k > k_{c}$ are linearly stable \citep{Sanz1994,Betti1995,Goncharov1996A}, which is in contrast to CRTI where the growth rate is a monotonically increasing function of $k$. Unlike the stabilization effect during the linear phase, the vortexes inside the bubble generated by mass ablation destabilize RTI in the nonlinear phase in both two-dimensional (2D) and 3D ARTI by the vortex acceleration mechanism \citep{Betti2006,Yan2016}. The vorticity provides a centrifugal force to the bubble vertex and accelerates the nonlinear terminal bubble velocity ($U_{b}^{rot}$) above the classical value ($U_{b}^{cl}$), especially for the small-scale 3D bubbles \citep{Betti2006,Yan2016}. An asymptotic analysis of ICF-relevant ARTI was performed and the nonlinear evolution of the front was analyzed in 2D geometry by a boundary integral method in the case of a strong temperature dependence of the thermal conductivity \citep{ALMARCHA2007}.

The large temperature gradient created by the ablation provides the well-known Nernst effect \citep{Nishiguchi1984} on the magnetic fields. The Nernst effect is known to provide an additional convective velocity against the direction of temperature gradient on the magnetic field, which will significantly affect the transport process of the magnetic field. One-dimensional (1D) simulations showed that the Nernst effect convects the magnetic field towards the high density region in laser-driven ablation plasma, and the magnetic field is significantly compressed and amplified \citep{Nishiguchi1984,Nishiguchi1985}.

The pioneering simulations \citep{Srinivasan2012A,Srinivasan2012B,Srinivasan2013} were performed on the magnetic field generation and evolution for 2D single-mode and multimode RTI in a stratified two-fluid plasma using a Hall MHD model. However, neither the mass ablation due to heat conduction nor the Nernst effect was considered in those simulations. Our previous simulations \citep{Zhang2022} including the heat conduction and the Nernst effect showed that $\sim$100 T magnetic fields can be generated via ARTI and the Nernst effect is a critical factor determining the magnetic fields' peak amplitude and spacial distribution. As feedback to hydrodynamics, the self-generated magnetic field changes electron thermal conduction by magnetizing the plasma. The analytical study of the effects of self-generated magnetic fields on ARTI in the linear regime \citep{ Garcia-Rubio2021} showed that the magnetic field affects the ARTI growth by bending the heat flux lines and it destabilizes ARTI for moderate Froude numbers ($Fr$) and stabilizes ARTI for large $Fr$, which is consistent with our 2D simulations \citep{Zhang2022}. The 2D simulations also showed that both the linear growth rate and the nonlinear amplitude of ARTI are increased by about $10\%$ due to the magnetic feedback \citep{Cui2024}. The simulations on the stagnation phase of an ICF implosion showed that the magnetic field can cool the spikes and weaken the ablative stabilization, which harmfully increases the heat loss of hot spot \citep{ Walsh2017}.

In this work, we present the simulation results for the evolution of the magnetic field generated via 3D ICF-relevant single-mode ARTI in a quasi-equilibrium frame of the acceleration phase of implosion. Important physics including ablation, Nernst, and magnetized heat conduction are taken into account to sketch more realistic magnetic fields' generation, evolution, and feedback to the ARTI evolution. It is found that $\sim 10^3$ T magnetic fields can be generated via 3D ARTI, which is an order of magnitude stronger than that found in our previous 2D work \citep{Zhang2022}. Such strong magnetic fields are able to profoundly alter local hydrodynamics by modifying the electron thermal conduction and speed up the growth of the spikes. The rest of the paper is organized as follows: section \ref{sec:model_setup} outlines the physical model and the simulation settings. In section \ref{sec:mag_evolution}, the simulation results on the magnetic fields' generation and transportation are presented and analyzed. In section \ref{sec:mag_heat}, the feedback of self-generated magnetic field on 3D ARTI nonlinear evolution is investigated. Section \ref{sec:summary} is a summary.

\section{Physics model and numerical method}\label{sec:model_setup}
The simulations on 3D ARTI and self-generated magnetic fields are carried out in planar geometry using the hydrodynamic code $\it{ART}$. Specifically  designed for modeling ARTI in ICF-relevant scenarios, \textit{ART} has been used in a number of recent works \citep{Betti2006,Yan2016,Zhang2018,Zhang2020,Xin2019,Xin2023,Zhang2022,LiuYang2023,Lijun2022,Lijun2023,LiJun2024,Fu2023}.
$\it{ART}$ solves the single-fluid equations in 2D/3D Cartesian coordinates with the Spitzer-Harm thermal conduction \citep{Spizter1953}. The hydrodynamic equations are as follows:
\begin{eqnarray}
	\label{eq:hydro_rho}
	\frac{\partial \rho}{\partial t} +\nabla \cdot (\rho \mathbf{v})=0,\\
	\label{eq:hydro_momentum}
	\frac{\partial\rho \mathbf{v}}{\partial t}+\nabla \cdot (\rho \mathbf{vv} )=-\nabla p +\rho\mathbf{g},\\
	\label{eq:hydro_energy}
	\frac{\partial \epsilon}{\partial t}+\nabla \cdot [(\epsilon+p)\mathbf{v}] =\rho\mathbf{v}\cdot\mathbf{g},\\
	\label{eq:thermal_conduction}
	\rho c_v \frac{\partial T}{\partial t}=\nabla\cdot(\kappa_{sh}\nabla T),
\end{eqnarray}
where $\rho$ is the mass density, $\mathbf{v}$ is the macroscopic single-fluid velocity of the plasma, $p$ is the plasma thermal pressure, and $\mathbf{g}$ is the acceleration. Since $\beta \gg 1$, the magnetic forces are neglected in the momentum equation.  The equation of state of an ideal gas is used, and the total energy is $\epsilon=[{p}/{(\gamma-1)}]+[{\rho v^2}/{2}]$ here. $T$ is the kinetic temperature including the Boltzmann constant, $\gamma={5}/{3}$ is the specific heat ratio, and $c_v$ is the constant-volume specific heat capacity. The single-temperature approximation of the plasma is applied so that the electron temperature $T_e$ and ion temperature $T_i$ are equal (ie. $T_e=T_i=T$).
The thermal conduction part (\ref{eq:thermal_conduction}) is solved separately from the energy equation (\ref{eq:hydro_energy}) in a Strang-splitting way \citep{Strang1968} and implicitly solved to avoid the strict time step $\Delta t$ requirement of explicit diffusion equation solvers. The thermal conductivity coefficient $\kappa_{sh}$ is given by the Spitzer-Harm model \citep{Spizter1953} without flux-limiter as $\kappa_{sh} \propto T^{5/2}$. A MUSCL-Hancock scheme \citep{MUSCL} with a HLLC \citep{HLLC} approximate Riemann solver is used as the hydrodynamic solver to achieve 2nd-order accuracy in both space and time. The single fluid is a DT plasma with the number ratio 1:1. The advantage of using a DT plasma in the simulations is that it can avoid complex physics such as radiation transport. Presently, neither radiation transport nor nonlocal electron heat conduction is included in our simulations of this paper.

The equation of magnetic field ($\mathbf{B}$)  evolution  can be readily derived from the Ampere's law, the Faraday's law, and the momentum equation of electrons \citep{Nishiguchi1984} and formulated in Gaussian units as
\begin{eqnarray}
\label{eq:magnetic}
\frac{\partial\mathbf{B}}{\partial t}=
\underbrace{\nabla\times\left(\mathbf{v}\times\mathbf{B}\right)}_{\rm{I}}
\underbrace{+\frac{c}{e}\nabla\times\left(\frac{\nabla p_e}{n_e}\right)}_{\rm{II}}
\underbrace{-\frac{c}{4\pi e}\nabla\times\left[\frac{(\nabla\times\mathbf{B})\times\mathbf{B}}{n_e}\right]}_{\rm{III}}
-\underbrace{\frac{c}{e}\nabla\times\frac{\mathbf{R}}{n_e}}_{\rm{IV}},
\end{eqnarray}
where $e$ is the elementary charge carried by an electron, $c$ is the speed of light in vacuum, $n_e$ and  $p_e$ are the number density and pressure of the electrons,  respectively. $\mathbf{R}=\mathbf{R_u}+\mathbf{R_T}$ as the transfer of momentum from ions to electrons caused by collisions consists of two parts: (i) the thermal force $\mathbf{R_T}$ due to the gradient of the electron temperatures and (ii) the friction force $\mathbf{R_u}$ due to the relative velocity of electrons and ions. $\mathbf{R_T}$ and $\mathbf{R_u}$ are given as follows \citep{Braginskii1965}:
\begin{eqnarray}
\label{eq:RT}
\mathbf{R_T}=-\beta_{\parallel}^{uT}\nabla_{\parallel}T_e
-\beta_{\perp}^{uT}\nabla_{\perp}T_e
-\beta_{\land}^{uT}\mathbf{b}\times\nabla T_e\\
\label{eq:RU}
\mathbf{R_u}=-\alpha_{\parallel}\mathbf{u}_{\parallel}
-\alpha_{\perp}\mathbf{u}_{\perp}
+\alpha_{\land}\mathbf{b}\times \mathbf{u},
\end{eqnarray}
where $\mathbf{b} \equiv \mathbf{B}/|\mathbf{B}|$ is the unit direction vector parallel to the magnetic field, $\mathbf{u} =\mathbf{u_e}-\mathbf{u_i} $ is the relative velocity of electrons and ions which can be associated with the magnetic field via the Ampere's law: $\mathbf{u}=-c\nabla\times\mathbf{B}/(4\pi n_ee)$. $\nabla_{\parallel}T_e$ and $\nabla_{\perp}T_e$ are the components of the temperature gradient parallel and perpendicular to the direction of $\mathbf{B}$, respectively. $\mathbf{u}_{\parallel}$ and $\mathbf{u}_{\perp}$ are the components of $\mathbf{u}$ parallel and perpendicular to the direction of $\mathbf{B}$, respectively.
Here, $\beta_{\parallel}^{uT}$, $\beta_{\perp}^{uT}$, $\beta_{\land}^{uT}$, $\alpha_{\parallel}$, $\alpha_{\perp}$, and $\alpha_{\land}$ are the plasma transport coefficients, which are formulated as
\begin{eqnarray}
	&\alpha_{\parallel}=\frac{m_en_e}{\tau_e}\alpha_0, \quad
	\alpha_{\perp}=\frac{m_en_e}{\tau_e}\left(1- \frac{\alpha_1'\chi^2+\alpha_0'}{\Delta} \right),\nonumber\\
	&\alpha_{\land}=\frac{m_en_e}{\tau_e}\frac{\chi(\alpha_1''\chi^2+\alpha_0'')}{\Delta}\nonumber\\
	&\beta_{\parallel}^{uT}=n_e\beta_0, \quad
	\beta_{\perp}^{uT}=n_e\frac{\beta_1'\chi^2+\beta_0'}{\Delta}, \nonumber\\
	&\beta_{\land}^{uT}=n_e \frac{\chi(\beta_1''\chi^2+\beta_0'')}{\Delta},\nonumber
\end{eqnarray}
respectively, where $m_e$ is electron mass, and the values of the coefficients $\alpha_0$, $\alpha_0'$, $\alpha_0''$,
$\alpha_1'$, $\alpha_1''$, $\beta_0$, $\beta_0'$ , $\beta_0''$, $\beta_1'$, $\beta_1''$, and ${\Delta}$ can be found in \cite{Braginskii1965}. Improved transport coefficients were later obtained \citep{Epperleinaa1986,Ji2023,Davies2021} through fitting numerical solutions of the Fokker-Planck equation. It was reported in \cite{Davies2021} that the fitted transport coefficients can give physically incorrect results under certain conditions. We still utilize the classical Braginskii's coefficients in this work, also to be consistent with our previous work on 2D ARTI\citep{Zhang2022} for a fair comparison.

Term I of (\ref{eq:magnetic}) is usually known as the convection term that freezes the magnetic field along with the plasma. Term II is the baroclinic term (also known as the Biermann battery) generating the self-magnetic field through the misaligned density and pressure gradients, since $c \nabla\times\left(\nabla p_e /n_e\right)/e=c\nabla p_e \times\nabla n_e /{(en_e^2)}$. Term III is neglected since the ratio of III to II approximately equals to  $1/\beta\ll1$ in the regimes covered by this work. Term IV brings the effects of collisions, including the magnetic dissipation related to $\mathbf{R_u}$ and the Nernst effect related to $\mathbf{R_T}$.
Bring in the expression of $\mathbf{R_T}$ and $\mathbf{R_u}$,  term IV  of (\ref{eq:magnetic})  can be written as
\begin{eqnarray}
	\label{eq:mag_nernst}
	&\frac{\partial \mathbf B}{\partial t}
	=\frac{c}{e}\nabla\times(\frac{\beta_{\parallel}^{uT}\nabla T_e }{n_e})
	-\frac{c}{e}\nabla\times[\frac{(\beta_{\parallel}^{uT}
	-\beta_{\perp}^{uT})(\mathbf{b} \times\nabla T_e)\times\mathbf{B}}{B n_e}] -\frac{c}{e}\nabla\times(\frac{\beta_{\land}^{uT}\nabla T_e\times \mathbf B }{ B n_e}) \nonumber\\
	&-\frac{c^2}{4\pi e^2}\nabla\times\{ \frac{\alpha_{\parallel}\nabla\times\mathbf{B}-\alpha_{\land}\mathbf{b}\times(\nabla\times\mathbf{B})-(\alpha_{\perp}-\alpha_{\parallel})\mathbf{b}\times[\mathbf{b}\times(\nabla\times\mathbf{B})]}{n_e^2} \},
\end{eqnarray}
The first term of the  right hand of (\ref{eq:mag_nernst}) has no contribution in a fully ionized plasma \citep{Sadler2021} while the combination of the second and third terms can be rewritten as
$\nabla\times[(\mathbf{V_N}+\mathbf{{V}_{CN}})\times\mathbf{B}]$, where $\mathbf{V_N}$ and $\mathbf{V_{CN}}$ are often referred as the Nernst velocity and the cross-gradient Nernst velocity in the form of
\begin{eqnarray}
\label{eq:vn}
\mathbf{V_N}=&-\frac{c \beta_{\land}^{uT}}{e B n_{e}}\nabla T_e,\\
\label{eq:vnc}
\mathbf{V_{CN}}=&-\frac{c(\beta_{\parallel}^{uT}-\beta_{\perp}^{uT})( \mathbf{b}\times\nabla T_e)}{e B n_e},
\end{eqnarray}
respectively. $\mathbf{V_N}$ is along the opposite direction of $\nabla T_e$ so that it convects the magnetic field in the direction of the heat flow. $\mathbf{V_{CN}}$ is in the direction of  $-\mathbf{b}\times\nabla T_e$, which causes the magnetic field to convect along the isothermal line. $V_{N}$ is much larger than $V_{CN}$ where $\chi \sim  {T_e^{3/2}B}/{n_e} \ll 1$, thus the contribution of $V_{N}$ is dominating with moderate magnetic fields. The last term of the  right hand of (\ref{eq:mag_nernst}) reflects the diffusion of the magnetic field in different directions.

The intense magnetic field changes the process of electron heat conduction by magnetizing the plasma.  While the magnetic field is not expected to be strong enough to significantly affect the implosion hydrodynamics via the momentum equation in ICF-relevant plasmas, it may be strong enough to magnetize the plasma and make the electron thermal conduction anisotropic via the Lorentz forces applied on the electrons moving in different directions with respect to the local magnetic field. The heat flux in a magnetized plasma reads as
\begin{equation}
	\mathbf{q_{mag}}=-\kappa_{\parallel}\nabla_{\parallel}T_e
	-\kappa_{\perp}\nabla_{\perp}T_e
	-\kappa_{\land}\mathbf{b}\times\nabla T_e,
	\label{eq:mag_q}
\end{equation}
where the detailed form of the magnetized plasma conduction coefficients $\kappa_{\parallel}$, $\kappa_{\perp}$ and $\kappa_{\land}$ can be found in  \cite{Braginskii1965}.
Along the direction parallel to $\mathbf{B}$, the magnetic field has no modification on the electron conduction and $\kappa_{\parallel}$ is identical to $\kappa_{sh}$. $\kappa_{\perp}$ is smaller than $\kappa_{\parallel}$ where $B\neq 0$ and $\kappa_{\perp}$ retreats to $\kappa_{sh}$ where $B=0$. $\kappa_{\perp}$ decreases as $\chi$ increases, leading to a flux-limiting effect perpendicular to the magnetic fields. The term $-\kappa_{\land} \bm{b}\times\nabla T_e$ known as the Righi-Leduc heat flux makes a special contribution to the heat conduction in a magnetized plasma. Both (\ref{eq:magnetic}) and (\ref{eq:mag_q}) have been implemented in \textit{ART} with the option to turn on and off different terms to be able to investigate a specific physical process.

\begin{figure}
	\centering
	\includegraphics[width=3.375in]{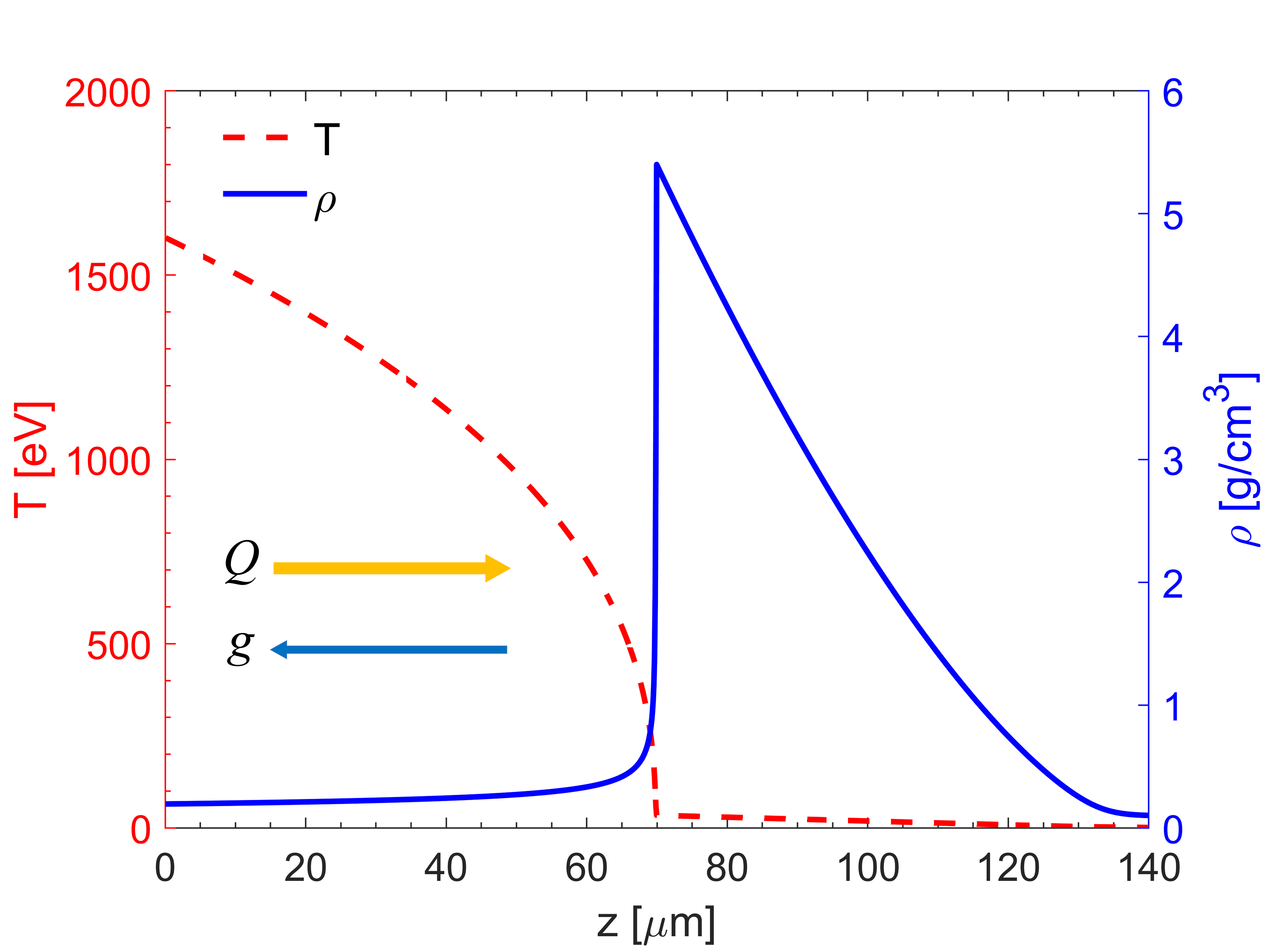}
	\caption{ The initial profile for $\rho$ (solid line) and $T$ (dashed line) along the z axis. }
	\label{fig:equibrium}
\end{figure}

In the \textit{ART} simulations, ARTI is initialized to grow from small perturbations on top of a quasi-equilibrium state abstracted from a typical profile of direct-drive NIF targets as shown in figure \ref{fig:equibrium}. The initial hydro profiles of a DT ablator are similar to those of a 1.5 MJ direct-drive ignition target \citep{McKenty2001} during the acceleration phase of an implosion. 
The cold and dense unablated DT shell is placed on top of the ablated DT plasma with a higher temperature but a lower density. The initial ablation front (the interface between the dense and the ablated plasma) is located at $z_0=70 \mathrm{\upmu m}$ with the peak density $\rho_a=5.3\mathrm{g/cm^3}$ reached on the ablation front. The quasi-equilibrium state  is obtained by integrating the 1D hydrodynamic equilibrium equations in the frame of reference of the shell from the ablation front toward both sides.

The ablation front is kept approximately fixed in space by balancing the ablative pressure with a dynamically adjusted effective gravity $g(t)$ which is initialized as ${g}(0)=100\upmu \mathrm{m/ns^2}$. Since the shell mass decreases due to ablation, the effective acceleration $g(t)$ is slowly and automatically adjusted in time during the simulation to keep the ablation front approximately fixed in space, i.e., $g(t) = [(p + \rho u^2)_{bot}-(p + \rho u^2)_{top}]/M_{tot}$, where the subscripts ``bot'' and ``top'' indicate the integral values at the bottom and top boundaries, respectively, and $M_{tot}$ is the total mass of the remaining plasma in the computational domain. This is equivalent to studying the ARTI growth in the frame of reference of the imploding shell. The quasi-equilibrium hydrodynamic profiles together with a time-dependent but spatially uniform gravity $g(t)$ are used to mimic an already well established quasi-equilibrium ICF plasma slowly evolving under the isobaric assumption also used in the analytical ARTI theories \citep{Goncharov1996A,Goncharov1996B}. Our simulations do not include the underdense region where the lasers are interacting with the plasma, and therefore we do not directly handle laser absorption. Instead, the laser energy transported toward the ablation front is simulated by a constant bottom-boundary heat flow calculated self-consistently with the SH model on the basis of the 1D hydrodynamic profiles to ablate the target at an ablation velocity $V_a = 3.5 \mathrm{\upmu \mathrm{m/ns}}$ (i.e., the penetration velocity of the ablation front into the heavy shell material), and the corresponding ablation pressure (i.e., the pressure at the ablation surface) is $p_a = 130 \mathrm{Mbar}$.

In order to seed 3D ARTI, the velocity perturbations ($\mathbf{v_p}$) are initialized around the ablation front in a divergence-free form as
\begin{eqnarray}
	&v_{px}=v_{p0} \sin(k x) \exp(-k\left\vert z-z_0 \right\vert), \nonumber\\
	&v_{py}=v_{p0} \sin(k y)\exp(-k\left\vert z-z_0 \right\vert), \nonumber\\
	&v_{pz}=v_{p0} [\cos(k x)+\cos(k y)]\exp(-k\left\vert z-z_0 \right\vert), \nonumber
\end{eqnarray}
where  $v_{p0}$ is the magnitude of the initial velocity perturbation set as $v_{p0} = 0.5\upmu \mathrm{m/ns}$, $k\equiv 2\pi/\lambda$ is the perturbation wavenumber, and $\lambda$ is the perturbation wavelength. A typical simulation is carried out with a simulation box of $140\upmu \mathrm{m}$ in the $z$ direction, while the widths in the $x$ and $y$ directions are chosen to be $\lambda$. A uniform Cartesian grid is used with the resolution of 10 grid points per $\rm 1 \upmu \mathrm{m}$ and the grid independence is checked to ensure numerical convergence. Periodic boundary conditions are applied in the $x$ and $y$ directions and the inflow/outflow boundary conditions are used on the upper/lower boundaries along the $z$ direction.

\section{Results and discussions}\label{sec:results}
A series of \textit{ART} simulations have been performed to study the generation, evolution, and feedback of the self-generated magnetic fields accompanying  ARTI in 3D geometry. As outlined in section \ref{sec:model_setup}, we have setup an idealized but still experimentally relevant scenario to be able to focus on the pure ARTI evolution and the magnetic generation in the \textit{ART} simulations. This approach also enables us to conveniently investigate the factors (i.e., geometric dimensions, ablation, the Nernst effects, and magnetized heat fluxes) that influence RTI and/or magnetic evolution by switching these modules in the simulations on and off while still allowing the simulations to start from virtually the same quasi-equilibrium hydrodynamic state, for a relatively fair comparison. The detailed simulation parameters spanning a range of $\lambda$, $V_a$, $p_a$, and $g_0$ are listed in table \ref{tab:param}. Our previous 2D work \citep{Zhang2022} demonstrated that the effects involving the self-generated magnetic fields are more profound for short-wavelength modes with $\lambda$ close to the linear cutoff wavelength $\lambda_c \equiv 2\pi/ k_c$. Therefore, the simulation cases are more focused on the short wavelength regime in this work.

\begin{table}
\begin{center}
\def~{\hphantom{0}}
\begin{tabular}{ccc ccc ccc}
	Case & 2D/3D & $\lambda(\upmu \mathrm{m})$ & $V_a(\upmu \mathrm{m/ns})$ & $p_a(\mathrm{Mbar})$ & $g_0(\upmu \mathrm{m/ns^2})$ & $\mathbf{V_{N}}$ & $\mathbf{V_{CN}}$ & $\mathbf{q_{mag}}$\\
	\hline
	\rmnum{1}  & 3D		& 10	& 3.5	& 140	& 100  	& off	& off 	& off  \\
	\rmnum{2}  & 3D		& 10	& 3.5	& 140 	& 100  	& on 	& on 	& off  \\
	\rmnum{3}  & 3D		& 10	& 3.5	& 140 	& 100  	& on 	& off 	& off  \\
	\rmnum{4}  & 2D		& 10	& 3.5	& 140 	& 100  	& on 	& on 	& off \\
	\rmnum{5}  & 3D		& 10	& 2.75	& 140 	& 100  	& on 	& on	& off \\
	\rmnum{6}  & 3D		& 10	& 2.0 	& 140 	& 100  	& on 	& on	& off \\
	\rmnum{7}  & 3D		& 10	& 3.5 	& 200 	& 100  	& on 	& on	& off \\
	\rmnum{8}  & 3D		& 10	& 3.5 	& 300 	& 100 	& on 	& on 	& off \\
	\rmnum{9}  & 3D		& 10	& 3.5 	& 140 	& 120 	& on 	& on 	& off \\
	\rmnum{10} & 3D		& 10	& 3.5 	& 140 	& 80   	& on 	& on	& off \\
	\rmnum{11} & 3D 	& 6 	& 3.5	& 140 	& 100   & on 	& on	& off \\
	\rmnum{12} & 3D 	& 20	& 3.5	& 140 	& 100   & on 	& on	& off \\
	\rmnum{13} & 3D 	& 30	& 3.5	& 140 	& 100   & on 	& on	& off \\
	\rmnum{14} & 3D 	& 6		& 3.5 	& 140 	& 100   & on 	& on 	& on  \\
	\rmnum{15} & 3D 	& 10	& 3.5 	& 140 	& 100   & on 	& on 	& on  \\
	\rmnum{16} & 3D 	& 20	& 3.5	& 140 	& 100   & on 	& on	& on  \\
	\rmnum{17} & 3D 	& 30	& 3.5	& 140 	& 100   & on 	& on	& on  \\
\end{tabular}
\caption{ Parameters  and physical  options of the \textit{ART} simulations}
\label{tab:param}
\end{center}
\end{table}

\subsection{Evolution of the self-generated magnetic field}\label{sec:mag_evolution}

\begin{figure}
	\centering
	\includegraphics[width=5.2in]{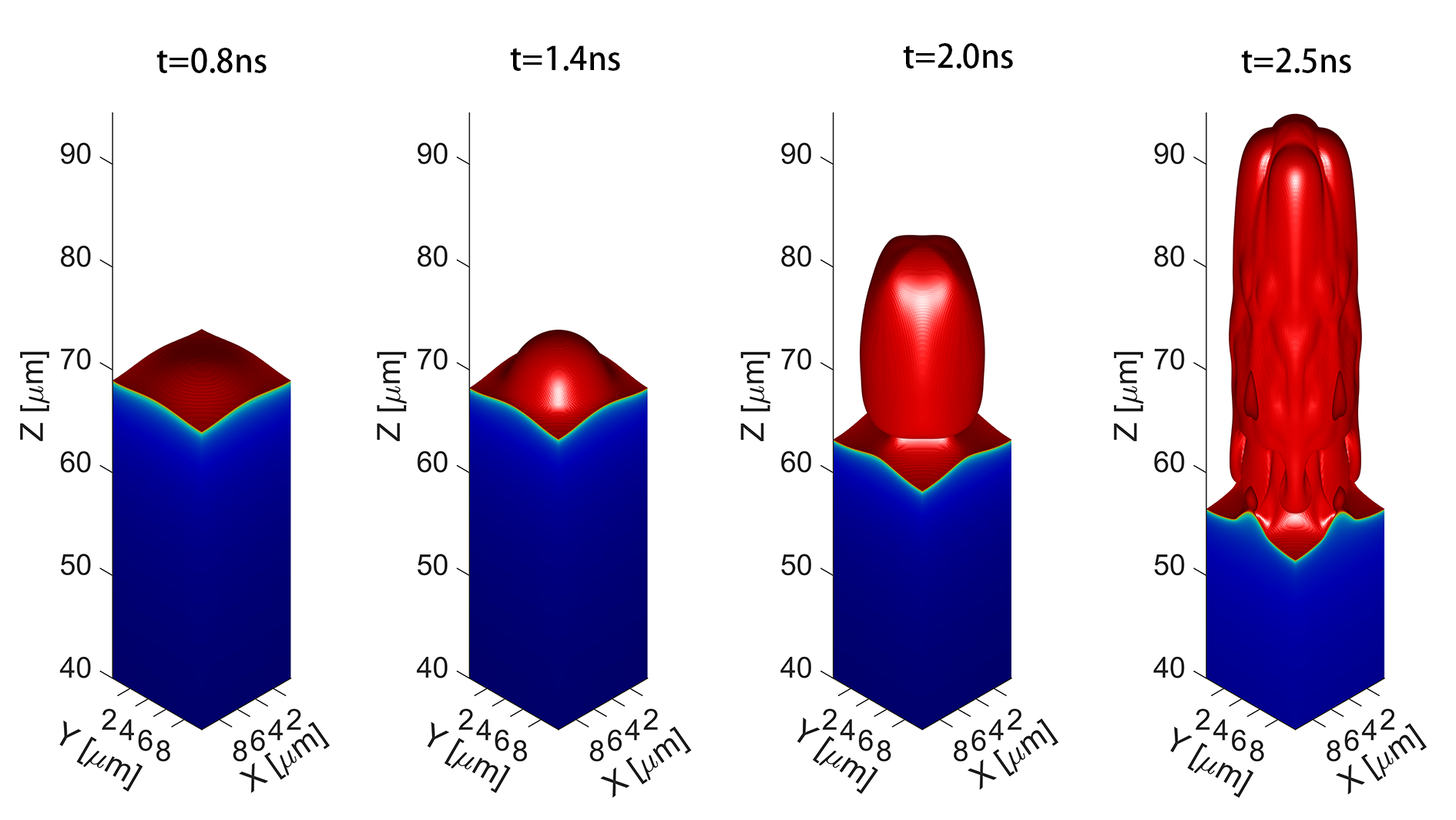}
	\caption{ The ARTI bubble structure at different times, from the linear stage [t = 0.8ns]  to the nonlinear stage [t = 1.4, 2.0 and 2.5ns], in Case i. }
	\label{fig:bubble_evolution}
\end{figure}

The evolution of ARTI from the linear stage up to the highly nonlinear stage in Case i is illustrated in figure \ref{fig:bubble_evolution}. It is shown that a clear 3D bubble is formed in the center of the simulation box as the perturbation amplitude increases, with the high-density fluid surrounding the bubble penetrating down into the low-density fluid. The ``bubble-spike'' topology is significantly different from its 2D counterpart.

\begin{figure}
	\centering
	\includegraphics[width=5.0in]{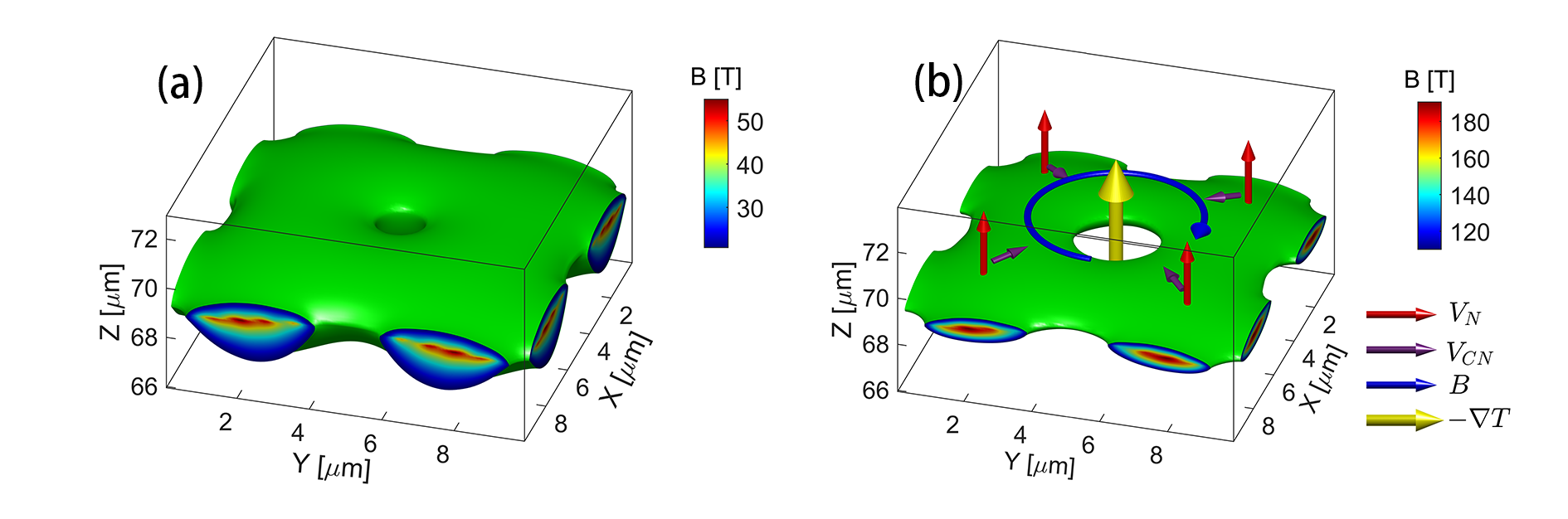}
	\caption{Magnetic field at the linear stage [t = 0.8 ns] of ARTI.  (a) Magnetic field isosurface at $B = 20 \mathrm{T}$ in Case i.  (b) Magnetic field isosurface of $B = 105 \mathrm{T}$  in Case ii, with the colored arrows illustrating the schematic of $\mathbf{V_N}$, $\mathbf{V_{CN}}$, $\mathbf{B}$, and $-\nabla T$ near the ablation front.}
	\label{fig:linear_mag}
\end{figure}

\begin{figure}
	\centering
	\includegraphics[width=5.2in]{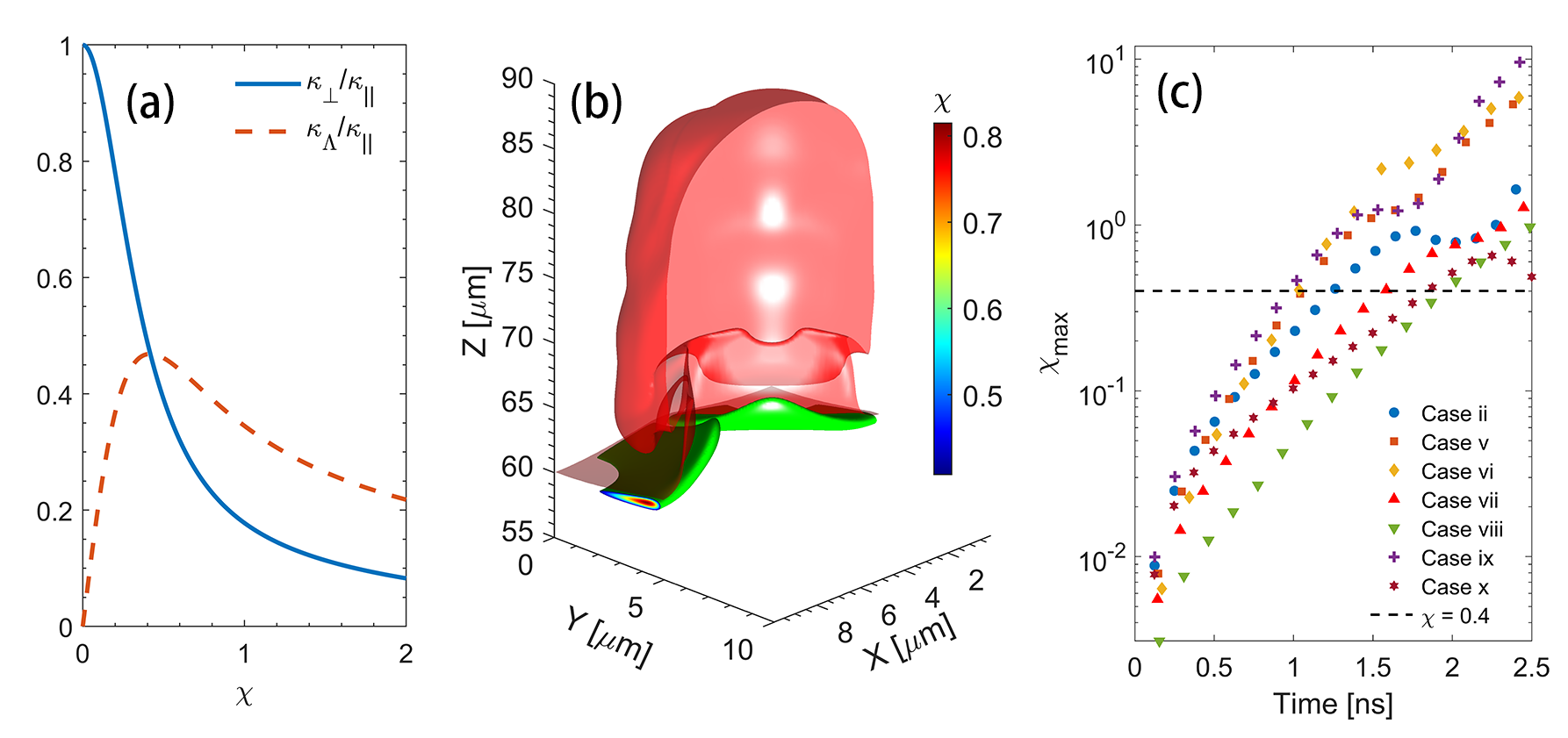}
	\caption{ Hall parameter in ARTI. (a) The ratios of $\kappa_{\perp}/\kappa_{\parallel}$ and  $\kappa_{\land}/\kappa_{\parallel}$ at different $\chi$ values. (b) Hall parameter at $t=2.0\mathrm{ns}$ in Case ii. The red surface is the bubble interface, and the green surface is the isosurface of the Hall parameter at $\chi=0.4$. (c) The peak value of Hall parameter $\chi_{max}$ vs. time in Cases ii and v-x, all with the Nernst effects included. }
	\label{fig:Hall_parameter}
\end{figure}

The magnetic field generation is found correlated with the growth of ARTI. The magnetic fields at the linear stage in Cases i (without Nernst) and ii (with Nernst) are illustrated in figure \ref{fig:linear_mag}(a) and (b) respectively. Figure \ref{fig:linear_mag}(a) shows that the magnetic fields are mostly generated near the ablation front where the baroclinic source ie. term II of (\ref{eq:magnetic}) is concentrated. The generated magnetic field then expands with the ablated material and enters the low-density area by convection. As a result, a magnetic field layer is formed below the ablation front. Case ii with the Nernst effects turned on clearly shows the influence from $\mathbf{V_{N}}$ and $\mathbf{V_{CN}}$ which are largely determined by the amplitude of the magnetic field and $\nabla T_e$ [see (\ref{eq:vn}) and (\ref{eq:vnc})]. The Nernst velocity $\mathbf{V_N}$ tends to convect the magnetic field against the ablation front [figure \ref{fig:linear_mag}(b)], where $T_e$ transits abruptly from the cold dense shell to the ablated low-density plasma thus $\mathbf{V_N}$ is towards the shell. The cross gradient Nernst velocity $\mathbf{V_{CN}}$ tends to transport the magnetic field towards the central axis of the bubble, illustrated by the arrows in figure \ref{fig:linear_mag}(b). ${V_{CN}}$ is read in the simulations to be much smaller than ${V_{N}}$ at the linear stage, consistent with the small Hall parameter at this moment [$\chi_{max} \ll 1$ as plotted in figure \ref{fig:Hall_parameter}(c)].  The Nernst effects lead to a stronger magnetic field in a thinner layer below the ablation front with the peak value of the magnetic field $B_{peak}$ amplified by more than three times.

\begin{figure}
	\centering
	\includegraphics[width=4.0in]{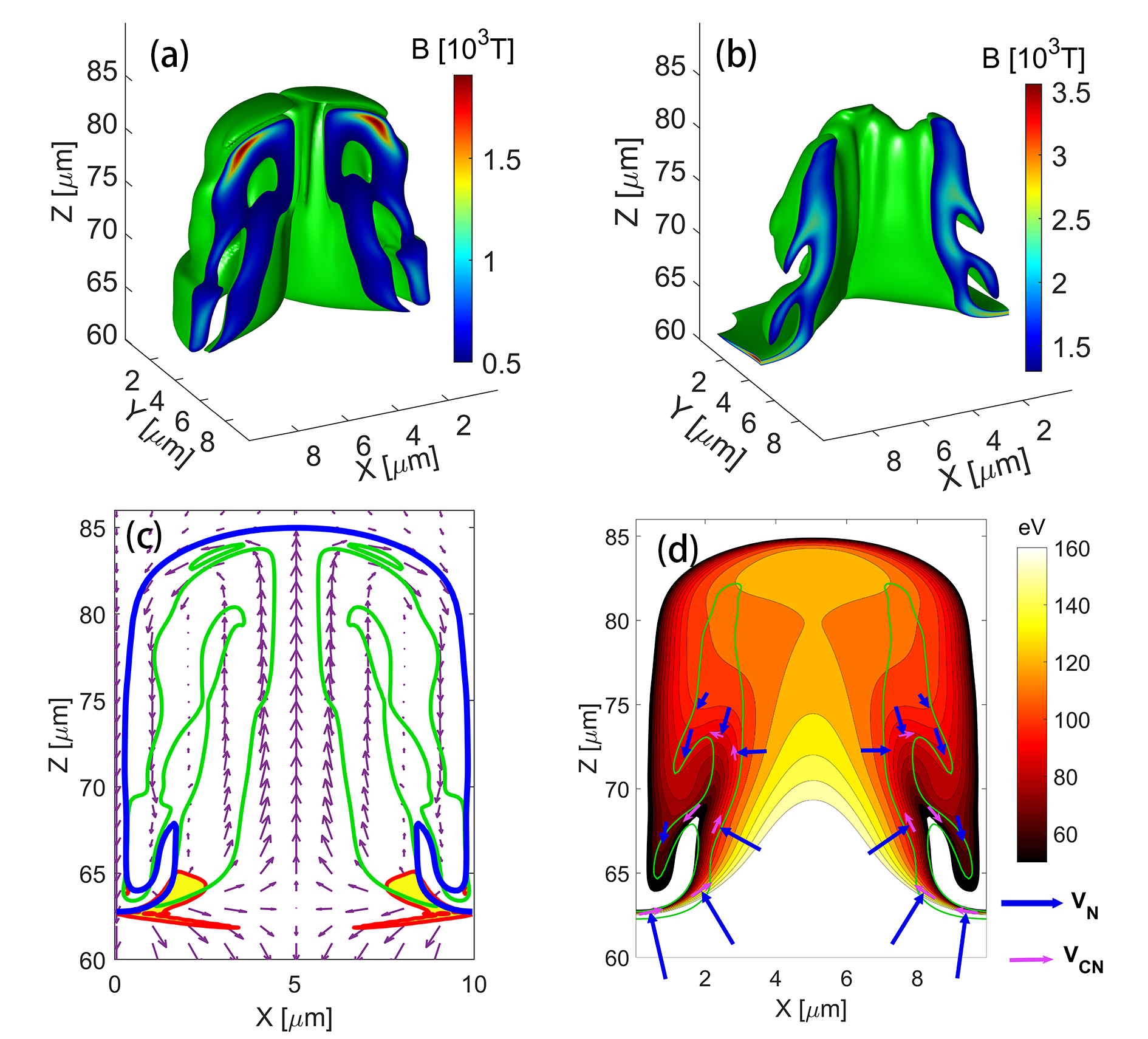}
	\caption{Comparison of self-generated magnetic fields at the highly nonlinear stage [$t = 2.0 \mathrm{ns}$] of 3D ARTI with and without the Nernst effect. (a) Magnetic field isosurface at $B = 500 \mathrm{T}$  in  Case i. (b) Magnetic field isosurface $B = 1400 \mathrm{T}$ in Case ii. (c)  Schematic diagram illustrating the transport of magnetic field with the plasma inside the bubble, taken at a slice passing through the bubble axis in Case i. Arrows denote fluid velocities relative to the bubble, defined as velocities of the light fluid within the bubble minus the penetration velocity of the bubble vertex into the denser fluid. The blue line represents the bubble interface, while the green lines indicate the magnetic field contours at $500 \mathrm{T}$. Yellow areas enclosed by red lines denote the regions where the baroclinic term exceeds $5000 \mathrm{T/ns}$. (d) Temperature contours inside the bubble at the slice passing through the bubble axis in Case ii. The green lines are the magnetic field contours at $1400 \mathrm{T}$. The arrows of different colors depict the schematic diagrams of $\mathbf{V_N}$ and $\mathbf{V_{CN}}$ inside the bubble. }
	\label{fig:nonlinear_mag}
\end{figure}

It is usually considered that ARTI enters the nonlinear stage when the perturbation amplitude grows larger than 0.1 $\lambda$ and a ``bubble'' emerges. In the nonlinear stage the growth of the ARTI bubble is subject to the vortex dynamics near the ablation front and inside the bubble. The magnetic fields in Cases i (without Nernst) and ii (with Nernst) are shown in figure \ref{fig:nonlinear_mag} for comparison. Figure \ref{fig:nonlinear_mag}(a) shows a complicated magnetic ring inside the bubble in Case i  while a slice passing the bubble axis is plotted in figure \ref{fig:nonlinear_mag}(c) to illustrate the transportation of the magnetic field. As the large-amplitude bubble is formed, the fluid inside the bubble gets cooled down [see figure \ref{fig:nonlinear_mag}(d)] thus the ablation on the interface inside the bubble is almost inhibited. Intense ablation is mostly concentrated on the ``spike'' tips where the baroclinic source generating magnetic fields also reaches the maximum as shown by the yellow areas in figure \ref{fig:nonlinear_mag}(c). Then the magnetic field is transported into the bubble along with the fluid convection. The arrows demonstrate the fluid velocities relative to the bubble motion, namely how the ablated light fluid carrying the magnetic field moves around inside the bubble. The ablated light fluid first moves upward toward the bubble vertex then downward guided by the bubble wall, forming a fairly complicated magnetic ring. The magnetic field reaches the maximum at the top of the bubble in Case i, as shown in figure \ref{fig:nonlinear_mag}(a).

Due to the similarity of the equations on the magnetic field evolution (\ref{eq:magnetic}) and on the vortex dynamics, the evolution of the magnetic field without Nernst effect  is quite similar to the evolution of vorticity $\mathbf{\upomega}  \equiv \nabla\times \mathbf{v}$.  (\ref{eq:magnetic}) retreats to $\partial_t\mathbf{B}=\nabla\times\left(\mathbf{v}\times\mathbf{B}\right) +({c m_i}/{2e})\nabla\times\left({\nabla p}/{\rho}\right)$ for a DT plasma if not considering the resistivity or Nernst effect, where $m_i$ is the average ion mass, while $\partial_t\mathbf{\upomega}=\nabla\times\left(\mathbf{v}\times\mathbf{\upomega}\right)-\nabla\times\left({\nabla p}/{\rho}\right)$ describes the vorticity evolution in a non-viscous fluid with conservative body force.  The self-generated magnetic field can be considered as an approximate signature of the vorticity in a non-Nernst fluid since $\mathbf{B}\approx -(cm_i/2e)\mathbf{\upomega}$, which is verified in the simulation of Case i.

The Nernst effects are found to alter the magnetic field distribution in the nonlinear stage of ARTI in Case ii, as shown in figure \ref{fig:nonlinear_mag}(b) and (d). Compared to its non-Nernst counterpart in Case i, the magnetic fields are more concentrated towards the spike tips and reach much higher magnitudes by roughly 2 times. The peak value of the magnetic field $B_{peak}$ with the Nernst effect is approximately $3.5 \mathrm{kT}$, whereas $B_{peak}$ without the Nernst effect is around $1.8 \mathrm{kT}$. Figure \ref{fig:nonlinear_mag}(d) shows the slice passing the bubble axis to illustrate the temperature distribution inside the bubble, which largely determines $\mathbf{V_{N}}$ and $\mathbf{V_{CN}}$. The schematic on the directions of $\mathbf{V_{N}}$ and $\mathbf{V_{CN}}$ is also demonstrated with the arrows. The fluid inside the bubble gets rapidly cooled down once leaving the spike tips, forming an intense temperature gradient and a very large  $\mathbf{V_{N}}$ inside the bubble towards the spike tips. So the magnetic field is  compressed by the $\mathbf{V_{N}}$ convection to a smaller area close to the spike tips compared to the non-Nernst Case i [figure \ref{fig:nonlinear_mag}(a)]. The cross gradient Nernst velocity $\mathbf{V_{CN}}$ tends to convect the magnetic field along the isotherm surfaces. However, as $V_{CN} \ll V_N$ is found in Case ii, the $\mathbf{V_{CN}}$ convection is expected to be less important than the  $\mathbf{V_{N}}$ convection, which is further verified by comparing to Case iii where $\mathbf{V_{CN}}$ is neglected. The magnetic distributions in Cases ii and iii are very similar.

\begin{figure}
	\centering
	\includegraphics[width=3.5in]{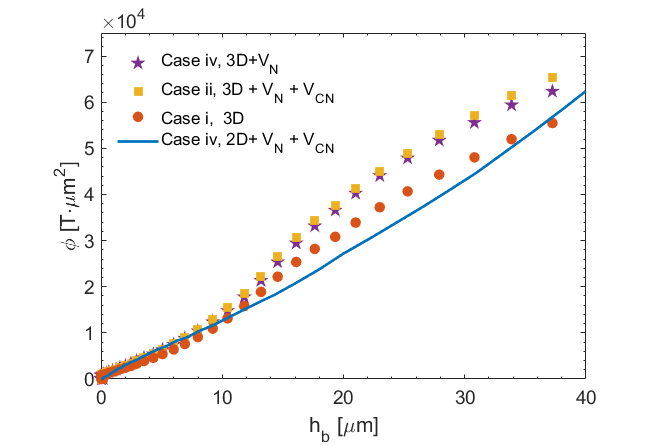}
	\caption{ The magnetic fluxes $\phi$ inside the bubble at different ARTI amplitudes ($h_b$), where $\phi$ is defined as  $\phi \equiv \int |B| dS$ at the slice of $y = \lambda/2$, and $h_b$ is defined as the height from the bubble vertex to the spike tip.}
	\label{fig:B_phi_nernst}
\end{figure}

Moreover, the growth of the  magnetic fluxes ($\phi \equiv \int |B| dS$ on an area passing the bubble axis and inside the bubble), which was used to evaluate an average intensity of the magnetic field \citep{Zhang2022},
with the ARTI bubble amplitude $h_b$  throughout a series of 3D and 2D simulations (Cases i-iv) with/without Nernst effects are plotted in figure \ref{fig:B_phi_nernst}. Cases i-iv use the same perturbation wavelength $\lambda = 10 \mathrm{\upmu m}$. It was found that $\phi$ inside 2D ARTI bubbles  are  monotonically correlated with  $h_b$ in our previous research, and can be well formulated by a scaling law, no matter if the Nernst effects are included \citep{Zhang2022}. It was also found that the Nernst velocity affects the convection process but not the generation of magnetic field and has little impact on $\phi$ in the 2D cases. Figure \ref{fig:B_phi_nernst} shows that $\phi$ in the 3D cases have similar behaviors to that in the 2D cases, as $\phi$ increases monotonically with $h_b$. While the peak value of the 3D magnetic field (about 3.5kT) is much larger than that of 2D (about 1.5kT), 3D $\phi$ (Cases ii and iii) is just slightly larger than 2D $\phi$ (Case iv) at the same $h_b$. Among the 3D cases,
$\phi$ with the Nernst effects (Cases ii and iii) is just slightly larger than the non-Nernst $\phi$ (Case i), indicating that the Nernst effects have mild impact on $\phi$, which is qualitatively consistent with the findings in 2D reported in \cite{Zhang2022}. The small difference on $\phi$ between Case ii and Case iii is also evidencing that the influence of $\mathbf{V_{CN}}$ is insignificant.

\begin{figure}
	\centering
	\includegraphics[width=5.2in]{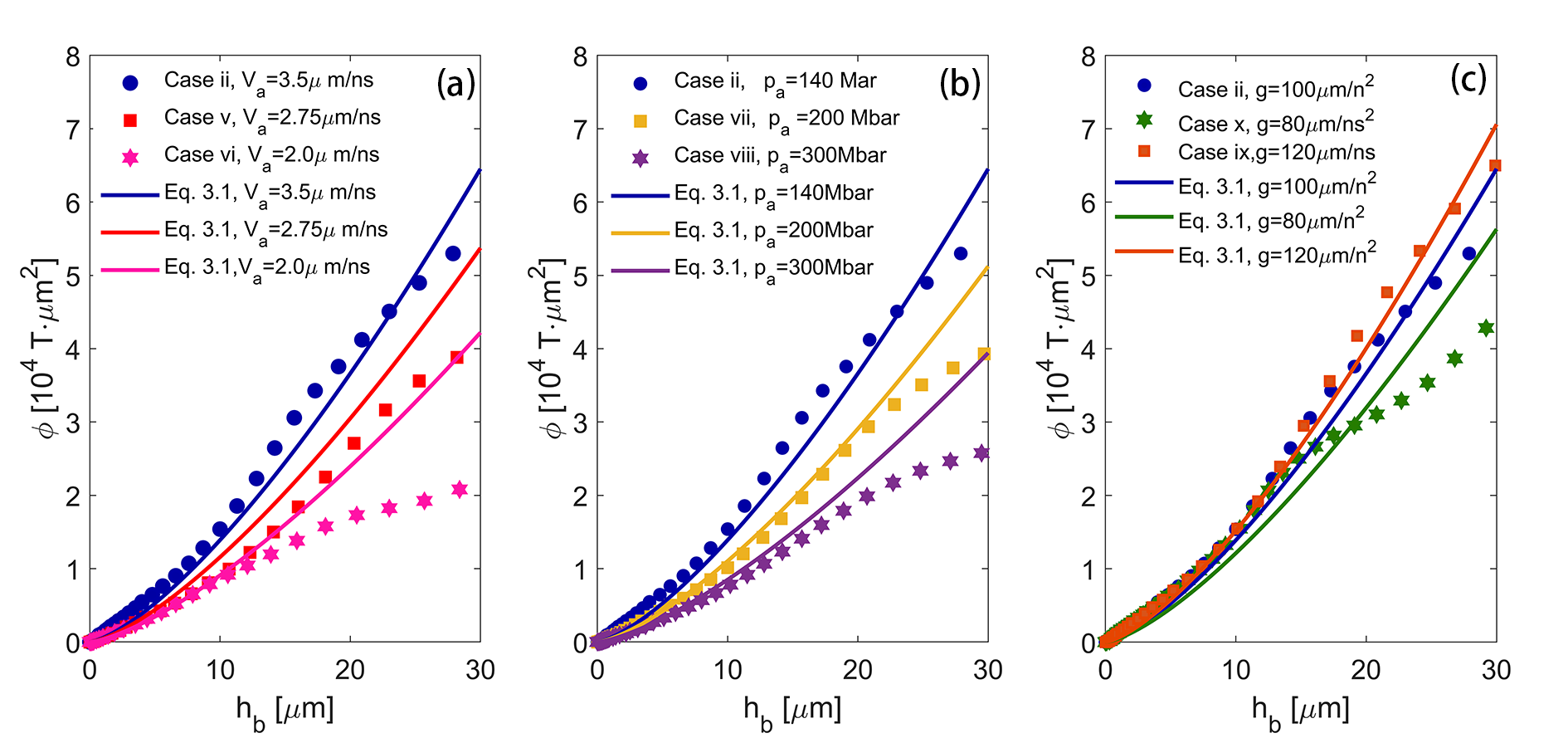}
	\caption{ The magnetic fluxes $\phi$ of $10 \upmu m$ simulations (Case ii,v-x). (a) The circles: Case ii. The squares: Case v. The stars: Case vi. (b) The circles: Case ii. The squares: Case vii. The stars: Case viii. (c) The circles: Case ii. The squares: Case ix. The stars: Case x. The solid lines in (a), (b), and (c) represent $\phi$ obtained by applying the magnetic flux scaling law (\ref{eq:scaling-law}) to the parameters of the cases with the same color, where $V_0 \equiv \sqrt{p_a/\rho_a}$, $g_0 \equiv p_a/(\rho_a \lambda)$, $\phi_0 \equiv c m_i \lambda \sqrt{p_a/\rho_a}/e$, $b=41.55$, $\xi=1.397$, $\eta=0.759$,  and $\theta=0.267$.}
	\label{fig:phi_scaling_law}
\end{figure}

We then examine if the scaling law for $\phi$ obtained in the 2D cases \citep{Zhang2022} is applicable in the 3D cases. The scaling law reads
\begin{equation}
	\label{eq:scaling-law}
	\frac{\phi}{\phi_0}=b (\frac{h}{\lambda} )^{\xi} (\frac{V_a}{V_0} )^{\eta} (\frac{g}{g_0} )^{\theta},
\end{equation}
where $V_0 \equiv \sqrt{p_a/\rho_a}$, $g_0 \equiv p_a/(\rho_a \lambda)$, and $\phi_0 \equiv c m_i \lambda \sqrt{p_a/\rho_a}/e$. The coefficients $b=41.55$, $\xi=1.397$,  $\eta=0.759$ and $\theta=0.267$ were fitted using the 2D simulation data. The scaling law (\ref{eq:scaling-law}) demonstrates positive correlations between $\phi$ and $V_a$ and $g$, and a negative correlation between $\phi$ and $p_a$. Figure \ref{fig:phi_scaling_law}(a), (b), and (c) plot the curves of $\phi$ versus $h_b$ in Cases ii and v-x at different values of $V_a$, $p_a$, and $g$, respectively. It is shown that larger $V_a$ or $g$ lead to larger $\phi$ at the same $h_b$ in figure \ref{fig:phi_scaling_law}(a) and (c), while figure \ref{fig:phi_scaling_law}(b) shows that smaller $p_a$ leads to a faster $\phi$ growth with $h_b$. These trends are consistent with the prediction of the scaling law (\ref{eq:scaling-law}). Moreover, figure \ref{fig:phi_scaling_law} shows that the 2D scaling law works reasonably well predicting $\phi$ in the 3D cases when $h_b$ is small. Larger deviations between the simulations and the predictions of the scaling law (\ref{eq:scaling-law}) show up as $h_b$ becomes larger, which is likely due to the 3D effects including the more complicated 3D magnetic structures inside the bubble.

As the magnitude of the magnetic field increases with $h_b$, the modification on the heat conduction is expected to be more significant. The heat flux perpendicular to the  magnetic field $\mathbf{q_{\perp}} \equiv -\kappa_{\perp}\nabla_{\perp}T_e$ is mitigated while the  Reghi-Leduc heat flux $\mathbf{q_{RL}} \equiv -\kappa_{\land}\mathbf{b}\times\nabla T_e$ that was absent in a magnetic-free plasma shows up to contribute to heat conduction.
The significance of the magnetic modifications on the heat conduction is usually evaluated by the Hall parameter $\chi$. Figure \ref{fig:Hall_parameter}(a) shows the dependence of the ratios of $\kappa_\perp$ and $\kappa_\land$ to $\kappa_\parallel$ on $\chi$ in a DT plasma.
It is shown that $\kappa_\perp/\kappa_\parallel$  decreases monotonically with the increase of $\chi$, while $\kappa_\land/\kappa_\parallel$ reaches the maximum at $\chi \approx 0.4$ where $\kappa_\perp/\kappa_\parallel$ drops by about a half. It is convenient to define a characteristic $\chi_c = 0.4$ such that the feedback of magnetic field to the heat conduction is considered to be significant where $\chi$ approaches or even exceeds $\chi_c$.

Figure \ref{fig:Hall_parameter}(b) illustrates the isosurface (green) where $\chi = \chi_c$ of Case ii in the highly nonlinear stage at $t=2.0 \mathrm{ns}$. It shows that large $\chi$ is mainly concentrated near the spike tip, and the peak value of $\chi$ can be larger than 0.8. The maximum of $\chi$ in Cases ii and v-x with different physical parameters ($V_a$, $g$, and $p_a$) and all including the Nernst effects are plotted in figure \ref{fig:Hall_parameter}(c). It is shown that $\chi_c$ can be reached in all cases, which indicates that the feedback of the self-generated magnetic field on the thermal conduction and consequently on the hydrodynamics could be significant with ICF-relevant parameters. It was found that the magnetic field boosts the ARTI bubble velocities of the short-wavelength modes while has minimal effects on the long-wavelength modes in 2D simulations \citep{Zhang2022}. The 3D simulations including the feedback are discussed in section \ref{sec:mag_heat} as follows.

\subsection{Effects of magnetized heat flux on the growth of 3D ARTI}\label{sec:mag_heat}
\begin{figure}
	\centering
	\includegraphics[width=5.2in]{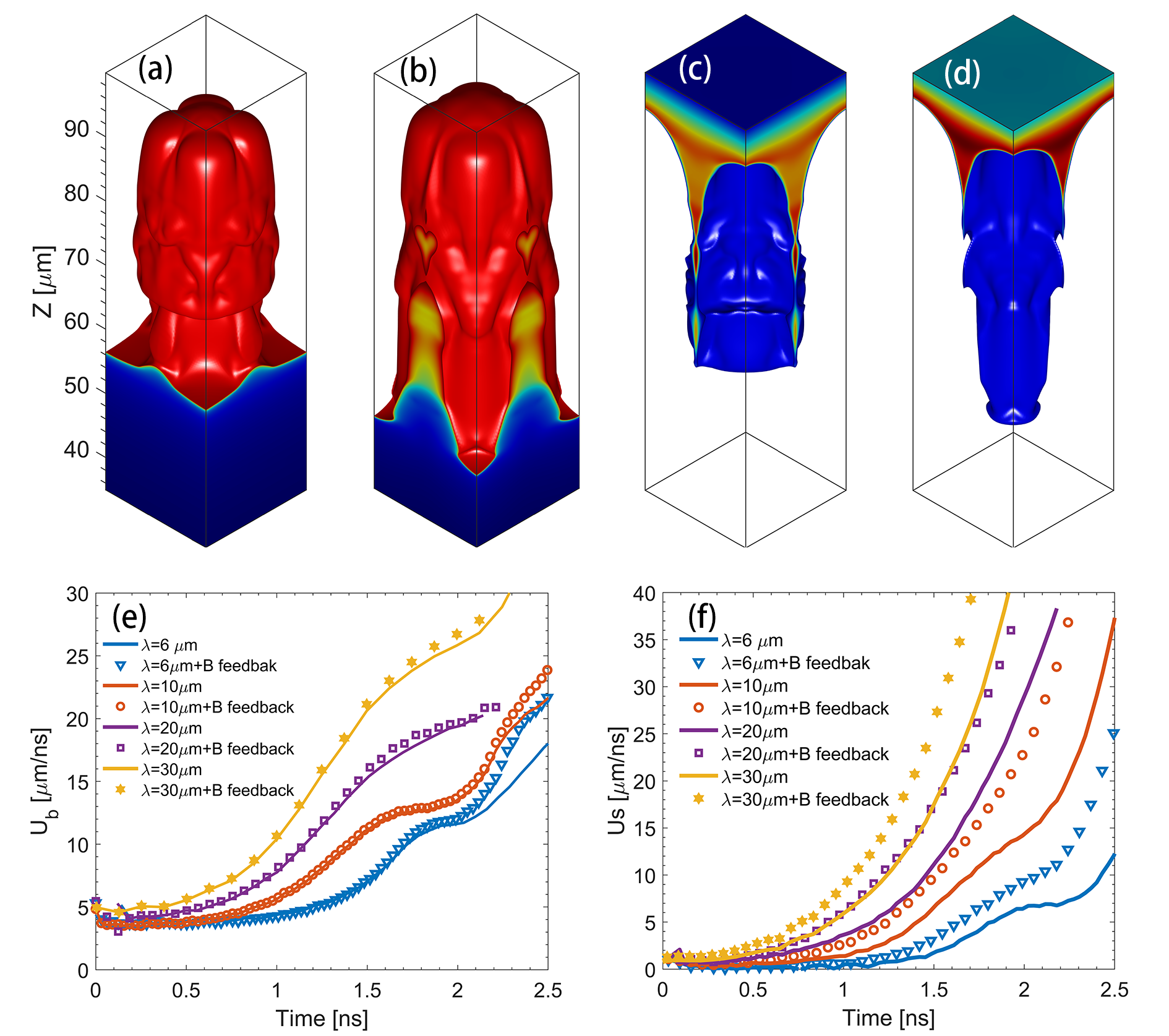}
	\caption{ (a-d) ARTI bubble and spike structures in the highly nonlinear stage [$t=2.5\mathrm{ns}$] for the case without [Case ii] and with [Case xv] magnetic feedback.  (a) and (c) are bubble and spike of the case without magnetized heat flux [Case ii]. (c) and (d) are bubble and spike of case with magnetized heat flux [Case xv].  The spike structures are plotted from data after the bubble translation to better demonstrate the effect of magnetic field feedback on spike growth. The bubble velocity $U_{b}$ (e) and the spike velocity $U_{s}$ (f) in the different wavelength cases without [solid lines]  and with [markers] magnetic feedback. The bubble velocity $U_b$ and the spike velocity $U_s$ are defined as the velocity of the vertex of the bubble and the spike relative to the velocity of the dense target plasma averaged in x-y plane.}
	\label{fig:bubble_spike}
\end{figure}

To investigate the feedback of the magnetic field on the hydrodynamic evolution in ARTI, a series of simulations [Cases xiv-xvii] using the magnetized heat flux described by (\ref{eq:mag_q}) are performed. The non-feedback cases [ii, xi-xiii] with the same other parameters but using the classical non-magnetized Spitzer-Harm heat flux are available for comparison. Four wavelengths [6, 10, 20 and $30 \upmu \mathrm{m}$] are selected in order to investigate the magnetic feedback at different wavelengths. The Nernst effects are always turned on in these simulation cases in this section to take into account the Nernst enhancement on the magnetic field as well as $\chi$.

The magnetized heat fluxes are found to significantly alter the ARTI bubble and spike  structures in the nonlinear stage. Figs. \ref{fig:bubble_spike}(a)/(c) and (b)/(d) compare the $\lambda = 10\upmu \mathrm{m}$ ARTI bubble/spike structures without [Case ii] and with [Case xv] the magnetized heat flux at the same time [$t=2.5 \mathrm{ns}$], respectively. It is shown that the magnetic feedback has little impact on the bubble evolution as the position of the bubble vertex with the magnetized heat flux [figure \ref{fig:bubble_spike}(b)] is only slightly higher than the non-feedback case [figure \ref{fig:bubble_spike}(a)]. However, the magnetized heat flux  has a significant impact on the growth and structure of the spikes, as shown in figure \ref{fig:bubble_spike}(c) and (d). The spike [figure \ref{fig:bubble_spike}(d)] with the magnetized heat flux is longer and finer than that [figure \ref{fig:bubble_spike}(c)] without the magnetized heat flux.

Moreover, the bubble velocity $U_b$ and the spike velocity $U_s$ without and with magnetic feedback are plotted in figure \ref{fig:bubble_spike}(e) and (f) for the simulations with different perturbation wavelengths to demonstrate the distinguished magnetic effects on the bubble and spike growths in 3D ARTI. It is shown that the magnetized heat flux has mild modifications on  $U_b$ in the  cases with $\lambda \geq 10 \upmu \mathrm{m}$, while just slightly increases $U_b$ in the  $\lambda = 6\upmu \mathrm{m}$ case in the highly nonlinear stage [$t>2.0 \mathrm{ns}$]. However, $U_s$ is increased more significantly than $U_b$ due to the magnetized heat flux in all the cases with different $\lambda$, as shown in figure \ref{fig:bubble_spike}(f). The behaviors of $U_b$ in the 3D cases are substantially different from the results in the 2D simulations \citep{Zhang2022} where the magnetized heat flux significantly increases the short-wavelength ($ \lambda \le 15\upmu \mathrm{m}$) $U_b$ in 2D. The magnetic modifications on $U_b$ in both 2D and 3D can be attributed to the ablation weakening near the spike tip.

\begin{figure}
	\centering
	\includegraphics[width=5.2in]{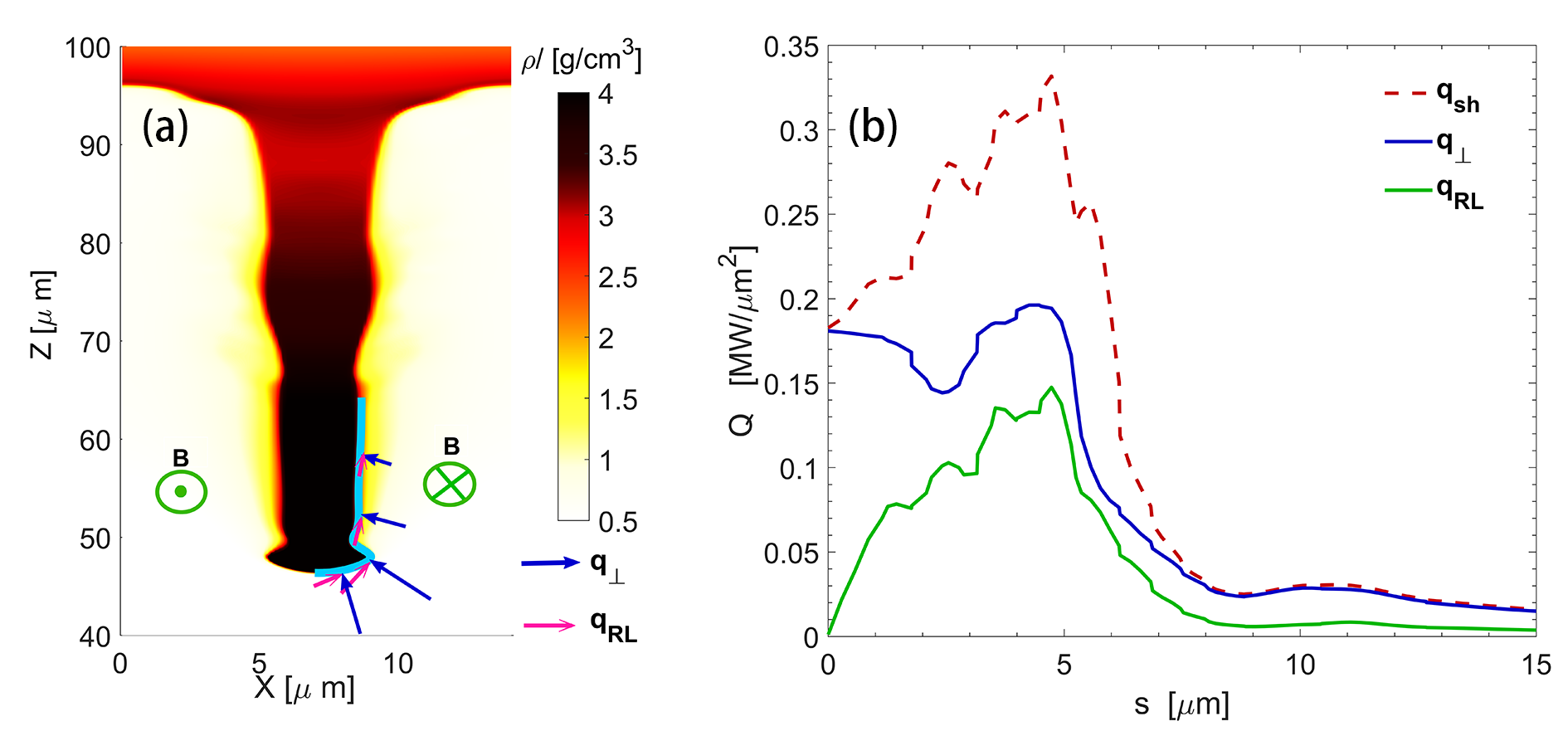}
	\caption{ (a) Density contour in the slice of $y=x$ and the schematic diagram of the magnetic field direction surrounding the spike in the nonlinear stage [$t=2.5 \mathrm{ns}$] for the $10\upmu \mathrm{m}$ wavelength case with magnetic feedback [Case xv]. The blue and pink arrows are the schematic diagram of $\mathbf{q_{\perp}}$ and $\mathbf{q_{RL}}$, respectively. (b) Heat flux at the spike interface [cyan curve in (a)] vs $s$, $s$ is defined as the distance on the cyan curve to the spike vertex. 
	}
	\label{fig:mag_q}
\end{figure}
The schematic of the magnetic fields, the magnetic-perpendicular heat flux $\mathbf{q_{\perp}}$, and the  Reghi-Leduc heat flux $\mathbf{q_{RL}}$ near a 3D spike are illustrated on top of the density contour in the highly nonlinear stage of Case xv in figure \ref{fig:mag_q} (a).  As $\nabla T_e$ is perpendicular to the magnetic field on the plane shown in figure \ref{fig:mag_q}(a), the heat flux $\mathbf{q_{\perp}}$ along $-\nabla T_e$ is reduced by the magnetic field, while the presence of $\mathbf{q_{RL}}$ tends to guide the heat flux  along the surface of the spike toward the inside of the bubble. Figure \ref{fig:mag_q}(b) further plots the heat flux components on the surface of the spike versus the distance $s$ on the cyan curve to the spike vertex. The unmagnetized classical Spitzer-Harm heat flux $q_{sh}$ is also plotted for comparison. It is shown that the strongest ablation is concentrated on the area near the spike tip as $q_{\perp}$ is large near the spike tip ($s<5\upmu \mathrm{m}$). $q_{\perp}$ is equivalent to $q_{sh}$ on the spike vertex ($s=0$) where it is magnetic-free. Elsewhere $q_{\perp}$ is significantly smaller than $q_{sh}$ near the spike tip, leading to weaker ablation on the spike. The Regi-Leduc heat flux whose peak value is comparable to $q_{\perp}$ tends to transport more heat into the bubble away from the spike vertex, which also helps to reduce the ablation near the spike tip. The weakened ablation helps to form the long spike in the magnetized-heat-flux Case xv as shown in figure \ref{fig:bubble_spike}(d). The different efficacy of the magnetic feedback boosting the bubble velocity in 2D and 3D is likely due to the different compressibility of the bubble in 2D and 3D. The width of a 2D bubble has to be squeezed as the spike gets wider due to the reduction of ablation and smaller bubble width is known to lead to stronger vorticity inside which supplies a stronger lifting force to the bubble vertex through the vortex acceleration mechanism \citep{Zhang2022,Betti2006}. However, a 3D bubble is more difficult to squeeze as it always tends to expand itself to form a round bubble. As shown in figure \ref{fig:bubble_spike}(a) and (b), the sizes of the upper part of the 3D bubbles in the cases with and without magnetic feedback are not notably different, which indicates that the vortex acceleration on a 3D bubble is not significantly enhanced by the magnetic feedback.

\begin{figure}
	\centering
	\includegraphics[width=5.2in]{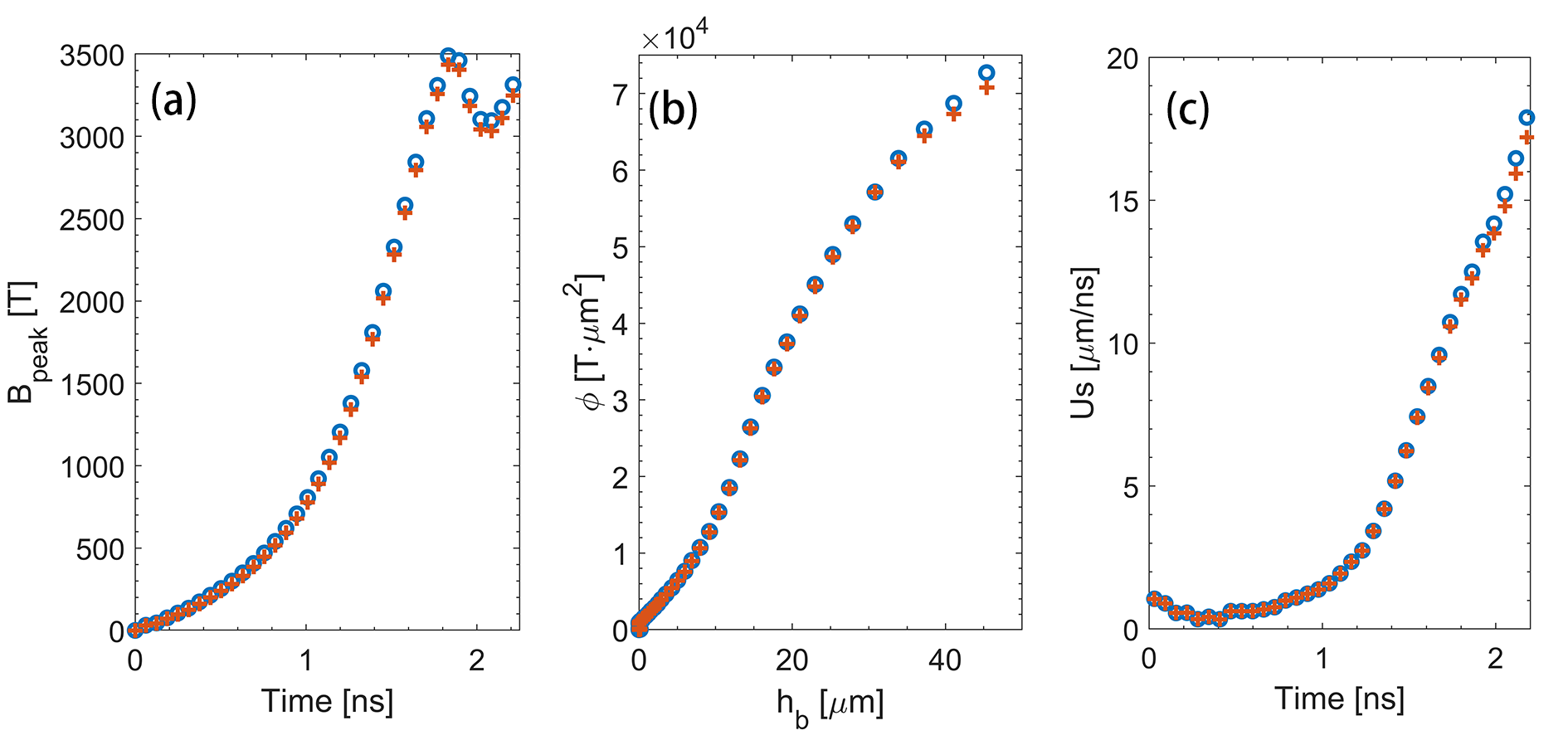}
	\caption{ $B_{peak}$ (a), $\phi$ (b) and $U_s$ (c) without (circles) and with (crosses) the magnetic field energy sink in the energy equation (\ref{eq:hydro_energy}). }
	\label{fig:B_energy_effect}
\end{figure}

In the cases presented so far (Cases i-xvii), the magnetic field energy have not been  coupled in (\ref{eq:hydro_energy}) in the large-$\beta$ regime where the magnetic field energy is negligible compared to the plasma internal energy.  However, the locally concentrated intense magnetic fields up to a few thousands of $\mathrm{T}$ generated in the highly nonlinear ARTI stages brings our attention on the validation of the modeling. We then take the magnetic field energy into account in the energy equation (\ref{eq:hydro_energy}) by putting the Biermann battery source to the magnetic field as an energy sink from the plasma internal energy. The simulation with the magnetic field energy sink are also performed, which has the same other simulation configurations as Case ii. The results of the simulation with the magnetic field energy sink and Case ii are plotted in figure \ref{fig:B_energy_effect}. It is shown that magnetic field energy sink has a very small influence on the $B_{peak}$, $\phi$, and $U_s$ in the highly nonlinear stage, indicating that the  magnetic field energy sink is unimportant in this regime.

\section{Summary}\label{sec:summary}
The self-generated magnetic field in 3D single-mode ARTI is numerically investigated with the parameters relevant to direct-drive ICF.  This study finds that 3D ARTI can produce much stronger magnetic fields, reaching magnitudes up to a few thousand T, compared to its 2D counterpart. Similar to the 2D cases, the inclusion of the Nernst effect significantly alters magnetic field convection and amplifies magnetic fields from the linear to the nonlinear stage of ARTI. The magnetic field is compressed toward the spike tip by the Nernst effect, reaching approximately twice of the peak magnitude in the non-Nernst case. Moreover, it is found that the influence coming from $\mathbf{V_{CN}}$ on the magnetic field convection is less significant than that from $\mathbf{V_{N}}$. The scaling law for the magnetic flux obtained from the 2D simulations performs reasonably well in the 3D cases, showing qualitatively similar dependence of the magnetic flux on  $V_a$, $p_a$, and $g$ in 2D and 3D.

In many 3D cases the Hall parameter can reach a characteristic value of  $\chi_c = 0.4$, beyond which the magnetized heat flux deviating significantly from the unmagnetized classical Spitzer-Harm heat flux is able to affect ARTI evolution. Unlike the  magnetic field significantly accelerates the bubble growth  in the  short-wavelength 2D modes, the magnetic field mostly accelerates the spike growth but has little influence on the bubble growth in 3D ARTI.  The spike acceleration due to magnetic feedback is attributed to the ablation reduction near the spike tip.

\begin{acknowledgments}
This research was supported by the National Natural Science Foundation of China (NSFC) under Grant Nos. 12175229 and 12388101, by the Strategic Priority Research Program of Chinese Academy of Sciences under Grant Nos. XDA25050400 and XDA25010200, by the Science Challenge Project, and by the Fundamental Research Funds for the Central Universities. The numerical calculations in this paper have been done on the supercomputing system in the Supercomputing Center of University of Science and Technology of China.
\end{acknowledgments}

Declaration of Interests. The authors report no conflict of interest.



\end{document}